\documentclass[iop]{emulateapj}
\usepackage{apjfonts}
\usepackage{graphicx}

\def\d3{$\delta_{3}$ }
\def\1d3{$(1 + \delta_{3})$ }
\def\l1d3{$\log_{10}(1 + \delta_{3})$ }

\def\s3{$\Sigma_{3}$}

\def\24m{24 $\mu$m}
\def\um{$\mu$m }

\def\sm{$M_{*}$}

\def\mpc{$h^{-1}$ Mpc}
\def\kms{${\rm km~s^{-1}}$ }

\def\Msolar{$\rm M_{\odot}$}

\def\deltaz{$\sigma_{\Delta~z/(1+z)}$}

\def\fq{$f_{q}$}
\def\lsim{\mathrel{\rlap{\lower4pt\hbox{\hskip1pt$\sim$}}
    \raise1pt\hbox{$<$}}}                
\def\gsim{\mathrel{\rlap{\lower4pt\hbox{\hskip1pt$\sim$}}
    \raise1pt\hbox{$>$}}}                

\shorttitle{Galaxy Environments in SPLASH}
\shortauthors{Lin et al.}

\begin{document}

\title{The SPLASH survey: Quiescent galaxies are more strongly clustered but are not necessarily located in high-density environments}

\author{Lihwai Lin \altaffilmark{1}, P. L. Capak \altaffilmark{2},  C. Laigle \altaffilmark{3}, O. Ilbert \altaffilmark{4}, Bau-Ching Hsieh \altaffilmark{1}, Hung-Yu Jian, \altaffilmark{1,5}, B. C. Lemaux, \altaffilmark{4}, J. D. Silverman \altaffilmark{6}, Jean Coupon \altaffilmark{7}, H. J. McCracken \altaffilmark{3}, G. Hasinger \altaffilmark{8}, O. Le F\'{e}vre, \altaffilmark{4}, N. Scoville \altaffilmark{9}}

\altaffiltext{1}{Institute of Astronomy \& Astrophysics, Academia Sinica, Taipei 106, Taiwan   (R.O.C.); Email: lihwailin@asiaa.sinica.edu.tw}
\altaffiltext{2}{Spitzer Science Center, California Institute of Technology, Pasadena, CA 91125, USA}
\altaffiltext{3}{Institut d'Astrophysique de Paris \& UPMC (UMR 7095), 98 bis boulevard Arago, F-75014 Paris, France}
\altaffiltext{4}{Aix Marseille Universit\'{e}, CNRS, LAM (Laboratoire d'Astrophysique de Marseille), UMR 7326, F-13388 Marseille, France}
\altaffiltext{5}{Department of Physics, National Taiwan University, 106, Taipei, Taiwan, Republic of China}
\altaffiltext{6}{Kavli Institute for the Physics and Mathematics of the Universe, Todai Institutes for Advanced Study, the University of Tokyo,
277-8583 Kashiwa, Japan}
\altaffiltext{7}{Astronomical Observatory of the University of Geneva, ch. d'Ecogia 16, CH-1290 Versoix, Switzerland}
\altaffiltext{8}{Institute for Astronomy, University of Hawaii, 2680 Woodlawn Drive, Honolulu, HI 96822, USA}
\altaffiltext{9}{California Institute of Technology, MC 249-17, 1200 East California Boulevard, Pasadena, CA 91125, USA}

\begin{abstract}

We use the stellar-mass-selected catalog from the $Spitzer$ Large Area Survey with Hyper-Suprime-Cam (SPLASH) in the COSMOS field to study the environments of galaxies via galaxy density and clustering analyses up to $z \sim 2.5$. The clustering strength of quiescent galaxies exceeds that of star-forming galaxies, implying that quiescent galaxies are preferentially located in more massive halos. When using local density measurement, we find a clear positive quiescent fraction--density relation at $z < 1$, consistent with earlier results. However, the quiescent fraction--density relation reverses its trend at intermediate redshifts ($1 < z < 1.5$) with marginal significance (<1.8$\sigma$) and is found to be scale dependent (1.6$\sigma$). The lower fraction of quiescent galaxies seen in large-scale dense environments, if confirmed to be true, may be associated with the fact that the star formation can be more easily sustained via cold stream accretion in `large-scale' high-density regions, preventing galaxies from permanent quenching. Finally, at $z>1.5$, the quiescent fraction depends little on the local density, even though clustering shows that quiescent galaxies are in more massive halos. We argue that at high redshift the typical halo size falls below $10^{13}$ \Msolar, where intrinsically the local density measurements are so varied that they do not trace the halo mass. Our results thus suggest that in the high-redshift Universe, halo mass may be the key in quenching the star formation in galaxies, rather than the conventionally measured galaxy density.

\end{abstract}


\section{INTRODUCTION}

It has long been recognized that star formation in galaxies is dependent on the local density of galaxies, usually referred to as the star formation rate (SFR)--density or color--density relation \citep{dre80,bal98,gom03,kau04,bal06,coo07,elb07}. These environmental trends are partly the result of a mass--density relation in which more massive galaxies are in general redder and older and are proportionally more abundant in denser environments \citep{kau04,bal06,muz12}. On the other hand, it has been shown that the SFR--density relation also holds at a fixed stellar mass \citep{kau04,pen10,lin14}, suggesting that environment is also a key factor in shaping the properties of galaxies. Furthermore, recent studies at $ z < 1$ have found that the effect of environment primarily alters the fraction of quiescent galaxies rather than reducing the SFR of galaxies on the star-forming main sequence \citep{muz12,koy13,lin14}, favoring a quenching process that operates over a short timescale (but see \citet{hai15} for an alternative perspective).  

While studies at $z$ > 1 have been undertaken, their is no clear consensus on the existence of a similar SFR--density relation at high redshift. While the environmental effect is apparently weaker at higher redshifts, it is debated whether the SFR--density relation flattens or even reverses beyond $z\sim1$ \citep{elb07,coo08,qua12,sco13}. The inconsistency among different results may be attributed to sample size, redshift uncertainties, and spectroscopic sampling rates \citep{coo10}, as well as varying definitions of environments -- physical environments (e.g., voids, field, filaments, clusters, etc.) versus local densities \citep{dar14}.  Recently, \citet{sco13} have measured the color--density relation in the COSMOS field out to $z\sim3$ 
in a self-consistent way and concluded that the environmental dependence is seen out to $z \sim 1.2$, but then becomes much weaker or none at higher redshifts, inconsistent with the trend predicted in the semi-analytic models in which the local color--density relation persists out to much higher redshits \citep{elb07,sco13}. 

Most environmental studies of galaxy properties beyond the local Universe are based either on a galaxy sample with a low (or moderate)  spectroscopic sampling rate or on a more 'complete' sample but with photometric redshifts. In both cases, the density measurements are subject to large uncertainties \citep{lai15}. Furthermore, the environmental dependence 
can be sensitive to the scale on which density is measured, because the mechanisms that transform galaxies from the star-forming sequence to the quiescent population may only be active on certain physical scales. Indeed, \citet{phl14} show that the red fraction--density relation has such a scale dependence and can be used to constrain the halo occupation distribution (HOD) prescription. Therefore, special attention is required when interpreting the color--density relation and its scale dependence.

Despite the fact that the local density field is widely used to probe environmental effects, it is not directly linked to the underlying large-scale structures, as evident by the large scatters in the relationship between local density and halo mass \citep{mul12}. Alternatively, galaxy clustering is an effective tracer of halo mass to investigate the environmental dependence of galaxy properties \citep{bau99,mo02}. Previous works on the clustering analysis of low-redshift galaxies have suggested that quiescent galaxies on average are located in more massive halos than star-forming galaxies at a fixed stellar mass \citep{li06,mos12}, in agreement with that inferred using local-density estimators. At redshifts greater than 1, the results based on a stellar-mass-selected sample are varied owing to the limited size of quiescent galaxy samples. For example, using $BzK$ color-selected $z \sim 2$ galaxies, \citet{lin12} found that quiescent galaxies have higher clustering amplitude than star-forming galaxies at a fixed stellar mass \citep[also see][]{sat14,mc15}, similar to the trend seen at lower redshifts. In contrast, \citet{bet14} study the $BzK$ galaxies selected from the COSMOS field and do not find a significant difference in the clustering strength between the quiescent and star-forming populations. 

In this study, we probe environmental effects on galaxy properties by combining two different approaches: local galaxy density and clustering, using data taken from the $Spitzer$ Large Area Survey with Hyper-Suprime-Cam (SPLASH) and the latest photometric redshift catalog in the COSMOS field \citep{laig15}. The SPLASH program consists of 2475 hrs ($>6$ hrs/pointing) of $Spitzer$ IRAC 3.6\um and 4.5\um~ observations of the two Hyper-Suprime-Cam Ultradeep fields, COSMOS \citep{sco07,san07} and
SXDS \citep{ued08}. The addition of the deep SPLASH data allows more accurate measurements of stellar mass for galaxies. Our paper is structured as follows. In \S2, we describe the sample and explain the redshift and stellar mass measurements. In \S3, we study the impact of photometric redshift uncertainty on the density measurement and the color--density relation. The main results are presented in \S4 and \S5. In \S6 we discuss the conclusions and implications of our results. Throughout this paper we adopt the following cosmology: \textit{H}$_0$ = 100$h$~\kms Mpc$^{-1}$, $\Omega_{\rm m} =
0.3$ and $\Omega_{\Lambda } = 0.7$. We adopt the Hubble constant $h$ = 0.7 when calculating rest-frame magnitudes. All magnitudes are given in the AB system.

\section{DATA AND SAMPLE SELECTIONS \label{sec:data}}

The COSMOS Ultra Deep Infrared Catalog \citep{laig15} includes 30 bands from UV to IR (0.25--7.7 \um). It contains the same optical dataset from previous releases \citep{cap07, ilb09}, the new $Y$-band data (PI: Guenther) taken with $Subaru$ Hyper-Suprime-Cam \citep{miy12}, the new near-infrared (NIR) available data from the UltraVISTA-DR2 \footnote{www.eso.org/sci/observing/phase3/data\_releases/uvista\_dr2.pdf}
 and IR data from the SPLASH program (P. L. Capak et al. in prep.). The photometry is extracted using SExtractor \citep{ber96} in dual image mode. Following a similar reduction procedure as described in \citet{mc12}, the detection image is a chi-squared combination of the four NIR images of UltraVISTA-DR2 ($Y,J,H,Ks$) and the optical band $z^{++}$ taken with $Subaru$ SuprimeCam. For regions that are not
covered by UltraVISTA, WIRCAM $H$ and $K$ data \citep{mc10} are used in this photometric catalog. The UltraVISTA
data include the DR2 `deep' and `ultra-deep' stripes, reaching $Ks$ $\sim$ 24.0 and 24.7 mag (AB, 3$\sigma$), respectively. Further details are described in \citet{laig15}. This work only utilizes data within the UltraVISTA footprint.

Following \citet{ilb13}, photometric redshifts are computed using LePhare \citep{arn02,ilb06}. Various templates are used, including spiral and elliptical galaxies from the libraries of \citet{pol07} and young blue star-forming galaxies from Bruzual and Charlot models \citep{bru03}. Predicted fluxes are computed in each photometric band for all redshift ranges between 0 and 6 with a step of 0.01. A correction for dust is included as a free parameter, using the Calzetti \citep{cal00} or Prevot (Prevot et al. 1984) extinction. The computation of the predicted fluxes takes into account the contribution of emission lines. Following the template-fitting procedure, a comparison of the photometric redshifts with the $z$COSMOS bright \citep{lil07} and faint (S. L. Lilly et al. in prep.) spectroscopic samples gives a precision of $\sigma_{\Delta_{z}/(1+z)}$ = 0.007 and an outlier rate $\eta$ = 0.5\% at $z<1.2$ ($i^{+} < 22.5$), and $\sigma_{\Delta_{z}/(1+z)}$ = 0.037 and $\eta$ = 9\% at $1.5 < z < 2.5$ ($i^{+} < 24$). Although it is not feasible to precisely quantify the photometric redshift performance for an NIR-selected sample as used in this work, we note that our simulation tests used to quantify the impact of photometric redshift error on the color--density relation extend to higher redshift uncertainties and outlier rates than quoted for the aforementioned $i$-selected samples. 

After obtaining an accurate photometric redshift, stellar masses are derived using the stellar population synthesis models of \citet{bru03}, assuming a Chabrier IMF \citep{cha03}, combining exponentially declining star formation history and delayed star formation history (SFR $\propto$ $\tau^{−2}e^{t}/\tau$) and considering solar and half-solar metallicities. 
Emission lines are added following the procedure of \citet{ilb09} by estimating the [OII] emission line flux from the UV luminosity of the rescaled template, using the \citet{ken98} calibration laws, and by adopting certain line ratios relative to [OII] for other lines. Two attenuation curves are considered, including the starburst curve of Calzetti et al. (2000) and a curve with a slope $\lambda^{0.9}$ (see the Appendix A of \citet{arn13}). $E(B−-V)$ values are allowed to vary up to 0.7.

\begin{figure}[h]

\includegraphics[angle=270,width=9cm]{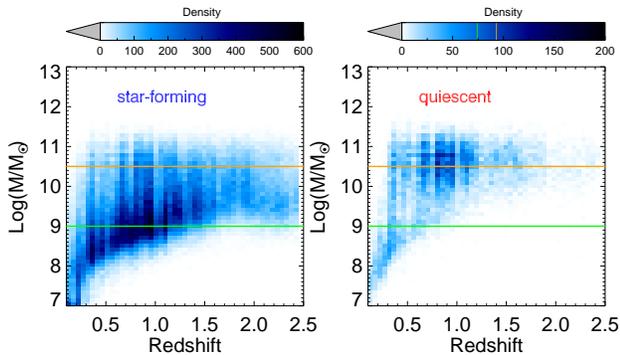}

\caption{Stellar mass as a function of redshift for galaxies with $K_{s} <24$ above 2$\sigma$ detections. The left (right) panel is for galaxies classified as star-forming (quiescent) population. The horizontal green lines denote the stellar mass limits above which are used for the local density calculation, whereas the orange lines correspond to the limits for computing the quiescent fraction, as well as for the clustering analysis.
\label{fig:sm-z}}
\end{figure}

\begin{figure*}

\includegraphics[angle=270,width=17cm]{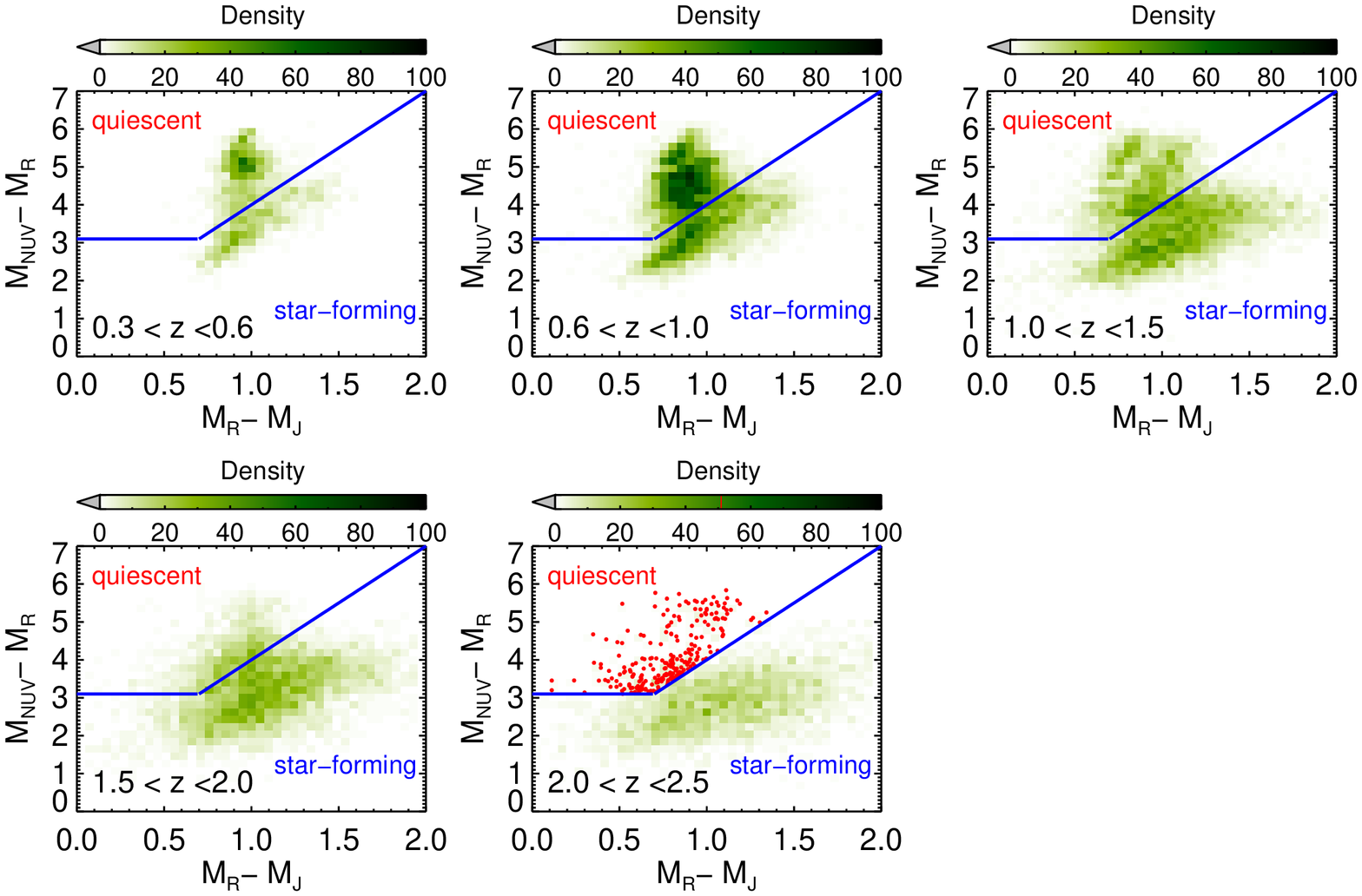}

\caption{Rest-frame $NUV-r^{+}$ versus $r^{+} - J$ colors for galaxies with \sm~ $> 10^{10.5}$ \Msolar. The two blue lines denote the boundary used to separate star-forming galaxies and quiescent galaxies. For clarity of the figure, individual quiescent galaxies are also shown as red dots in the highest-redshift bin ($2 < z < 2.5$). 
\label{fig:color}}
\end{figure*}

Since the NIR data are essential for the study of galaxies beyond $z \sim 1$, we first construct a faint galaxy sample having $Ks$ < 24, which roughly corresponds to the depth of the shallower part of UltraVISTA data, namely, the `deep' stripe. We further remove galaxies with signal-to-noise ratio < 2, as the $K$-band photometry for those galaxies does not provide useful information in terms of photometric redshift determination. For galaxies with \sm~ $ > 10^{9}$ \Msolar~ at $0.1 < z < 2.5$, all objects except four are detected in more than 25 filters, ensuring sufficient bandpasses for the computation of a photometric redshift and stellar population parameters. In Figure \ref{fig:sm-z}, we show the stellar mass versus redshift distribution for galaxies satisfying our $K$-band selection. Following \citet{ilb13}, we classify the star-forming and quiescent populations according to their locations on the NUV--$r^{+}$ and $r^{+}--J$ plane (see figure \ref{fig:color}). This color selection has been used to study the evolution of the stellar mass function for different types of galaxies up to $\sim 3$ \citep{ilb10,ilb13}. As can be seen in figure \ref{fig:sm-z}, the quiescent galaxies become progressively incomplete at high redshifts. \citet{laig15} estimate that the stellar mass limit for the quiescent galaxies is $\sim $ 10$^{10.3}$ \Msolar~ at $z \sim 3$.  We therefore further limit our analysis in \S4 and \S5 to galaxies with \sm~ > 10$^{10.5}$ \Msolar~ over the full redshift range of $0.3 < z < 2.5$. On the other hand, we include galaxies with \sm~ $ > 10^{9}$ \Msolar~ when computing the local density. One caveat of our density measurement is that the density field may be underestimated because of the incompleteness of quiescent galaxies in the range of 10$^{9}$ and 10$^{10.3}$ \Msolar~ at $z >1$. However, we note that the quiescent fraction decreases rapidly with decreasing stellar mass and is less than 20\% for galaxies with \sm~ $ \sim 10^{10}$ \Msolar~ even in the cluster environments at $z < 1$ \citep{lin14}. We therefore do not expect a significant impact on the density measurement owing to the lack of low-mass quiescent galaxies if they exist.

\section{DENSITY MEASUREMENTS \label{sec:density}}

The local density is computed using two common approaches used in the literature, the `fixed aperture' method  \citep[e.g.,][]{gal09,gru11}, in which we count the number of galaxies within a cylinder centered at the position of each galaxy, and the `$n^{th}$-nearest neighbor' method \citep[e.g.,][]{gom03,bal06,coo07}, for which the surface area is defined as the area enclosed by the $n^{th}$-nearest neighbor in the projected plane. The aperture size of the cylinder in the right ascension (RA) and declination (Dec) directions used when adopting the 'fixed aperture' method ranges from 0.2 to 0.9 physical \mpc. The length in the line-of-sight direction in both methods is set to be 0.04 $\times(1+z)$, comparable to the photometric redshift uncertainty of this catalog. \footnote{As tested in \citet{lai15}, varying the line-of-sight redshift window around the value corresponding to the photometric redshift uncertainty by a factor of 2 or so when computing the density field does not significantly change the results.} For each galaxy, the overdensity ($1+\delta$) is then defined as the density divided by the median density of all the galaxies located at similar redshifts (within $d_{z}$ of 0.04 $\times(1+z)$). Since the density is computed by searching for neighbors within a small redshift interval around each galaxy before binning the redshifts for the density and clustering analysis, the neighboring galaxies falling outside the edge of each redshit bin would still be counted for the density measurement.  Therefore, our results are not affected by the redshift edge effect that is mentioned in \citet{qua12}. 

\subsection{Impact of photometric redshift uncertainty on the density field}
\begin{figure*}
\includegraphics[angle=0,width=17cm]{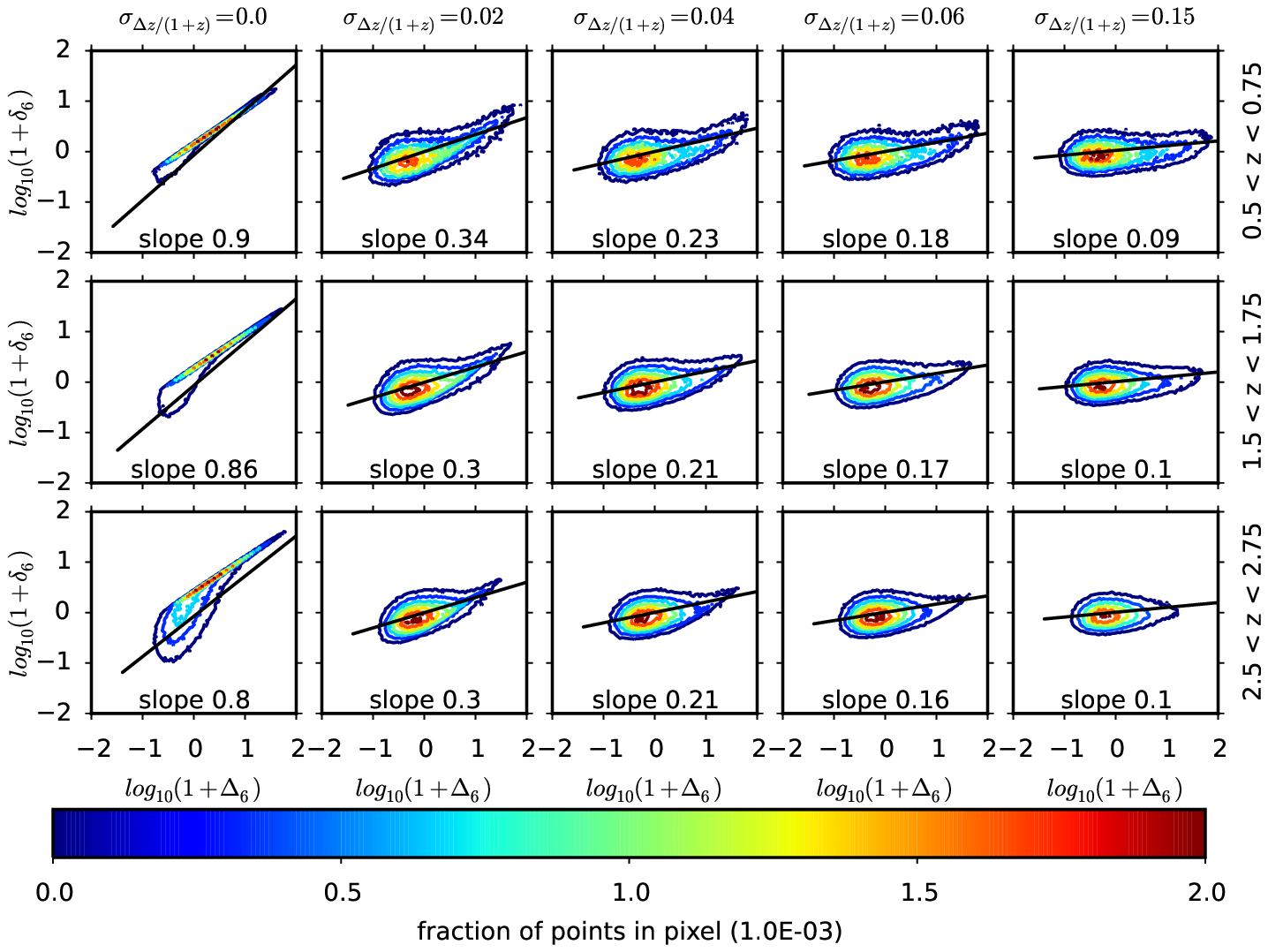}
\caption{Comparison between the surface density field ($y$-axis) and the 3D density field ($x$-axis) measured using the $6^{th}$-nearest neighbor method in a mock galaxy catalog. The solid lines correspond to the best fits to the contours. The photo-$z$ uncertainty increases from left to right, and the redshift range increases from top to bottom.
\label{fig:sigma}}
\end{figure*}

\begin{figure*}
\includegraphics[angle=0,width=17cm]{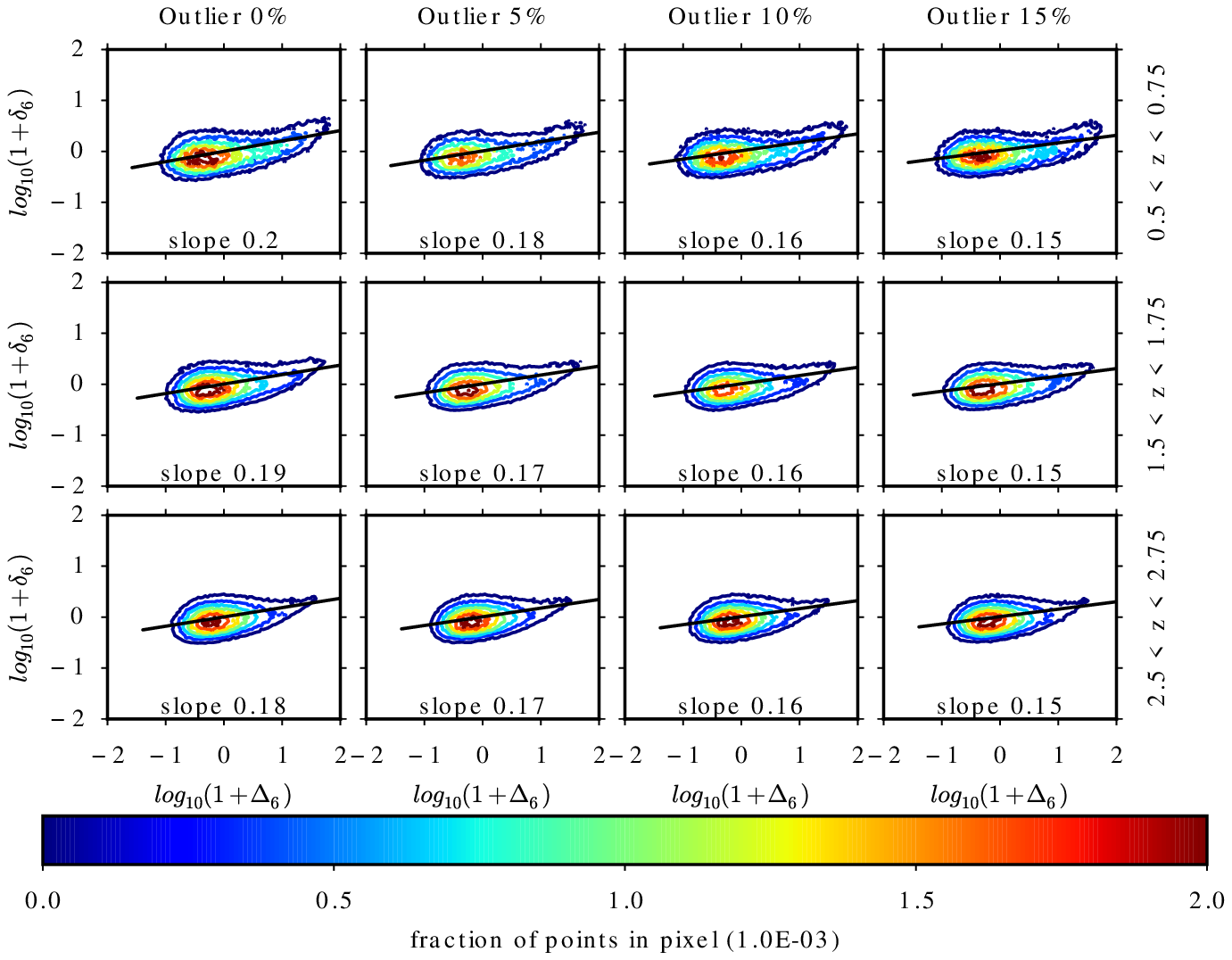}
\caption{Similar to figure \ref{fig:sigma}, but the results are shown by varying the outlier rate, increasing from left to right. The photo-$z$ accuracy is fixed to 0.05.
\label{fig:outlier}}
\end{figure*}

Because of the uncertainty on photometric redshifts, the local density measurement is likely subject to large errors. Therefore, it is crucial to quantify the reliability of the density measurements to investigate the correlation with galaxy properties. There have been many early works attempting to quantify the effect of photometric redshift uncertainly on the density determination. For example, \citet{coo05} conclude that the galaxy density measurement is problematic for samples with photometric redshift accuracy worse than 2\%. However, it was later realized that their work was based on a shallower magnitude-limited data set (in private communication with M. Cooper), and the simulation they used underestimates the density of galaxies at the cluster core \citep{cap07}. Subsequent studies, on the other hand, have demonstrated that photometric reshift datasets are promising for detecting the color--density relation by utilizing simulations \citep{cap07,qua12,lai15,dar15}. Recently, \citet{lai15} have studied the effects of photometric redshift precision and outliers in great detail and concluded that the color--density relation is detectable out to $z < 0.8$ even in the case of \deltaz~ up to 0.06. 

We followed a similar approach to the one described in \citet{lai15} to understand the systematics/bias of our density measurements and extend the analysis to $z \sim 3$ by using mock galaxy catalogs. The mock catalog was constructed by \citet{kit07} based on a semi-analytic galaxy formation model of \citet{cro06} as updated by \citet{del07}, implemented on the Millennium Simulation \citep{spr05} \footnote{The catalog is available on the Millennium download site for people who have registered an account (http://gavo.mpa-garching.mpg.de/Millennium/).}. 
This mock catalog covers a 1.4 $\times$ 1.4 deg$^{2}$ field and contains galaxies with stellar mass down to $10^{7}$ \Msolar~ over a wide redshift range (up to $z \sim 5$), which is suitable for our purpose. When computing the two-dimensional (2D) projected density field, we disturb the redshift of galaxies based on a Gaussian distribution with the mean equal to the true redshift of galaxies and width equal to a specified photometric redshift uncertainty.

Figures \ref{fig:sigma} and \ref{fig:outlier} show our 2D projected density measurements against the true three-dimensional (3D) local density as a function of redshift precision and outlier rate, respectively. Keeping in mind that the overdensity is defined in the 2D plane by projecting galaxies within a finite redshift interval, it is therefore expected that there would still exist some scatter in the density comparison even in the case of no redshift error, as shown in the three left panels of figure \ref{fig:sigma}.  As the photometric redshift accuracy worsens, the slope of this relation becomes shallower owing to the dilution of the density contrast. As it can be seen, the relation is nearly flat (slope $\sim$ 0.1) and the dynamical range of the 2D overdensity becomes quite small when the photometric redshift error increases to 15\%. 

In addition, we also study the impact of catastrophic failure of the photometric redshift determination (i.e., the redshift outlier) on the density measurement. For simplicity, we assign an arbitrary redshift to a certain fraction of randomly selected galaxies. In figure \ref{fig:outlier}, we show the 2D vs. 3D density estimates by varying the outlier rate. Overall, it is found that the resulting relationship between 2D and 3D measurements does not strongly depend on the outlier rate. Therefore, their effect can be ignored in this work.

\subsection{Impact of photometric redshift uncertainty on the color--density relation}
Since the photo-$z$ uncertainty yields a shallower slope in the 2D versus 3D density relation, one may expect that the color--density relation can potentially be smeared out if using the photometric redshfit sample. In the following, we make use of the mock galaxy catalog to examine this effect. For each galaxy in the mock catalog, we first compute their overdensity and then assign a galaxy to be quiescent or star-forming such that the fraction of quiescent galaxies obeys the following formula:
\begin{equation}\label{eq:fq}
f_{q} = 0.5 + 0.1 \times log_{10} (1 + \delta_{6}),
\end{equation}
where (1 + $\delta_{6}$) is the overdensity measured using the $6^{th}$- nearest neighbor. The above relation between quiescent fraction and overdensity is arbitrarily set and hence may not be representative of the underlying color--density relation predicted by the semi-analytical model used in the mock catalog. Nevertheless, this simplified color--density relation should be sufficient for our purpose of understanding the effect of redshift error.

\begin{figure*}
\includegraphics[angle=0,width=17cm]{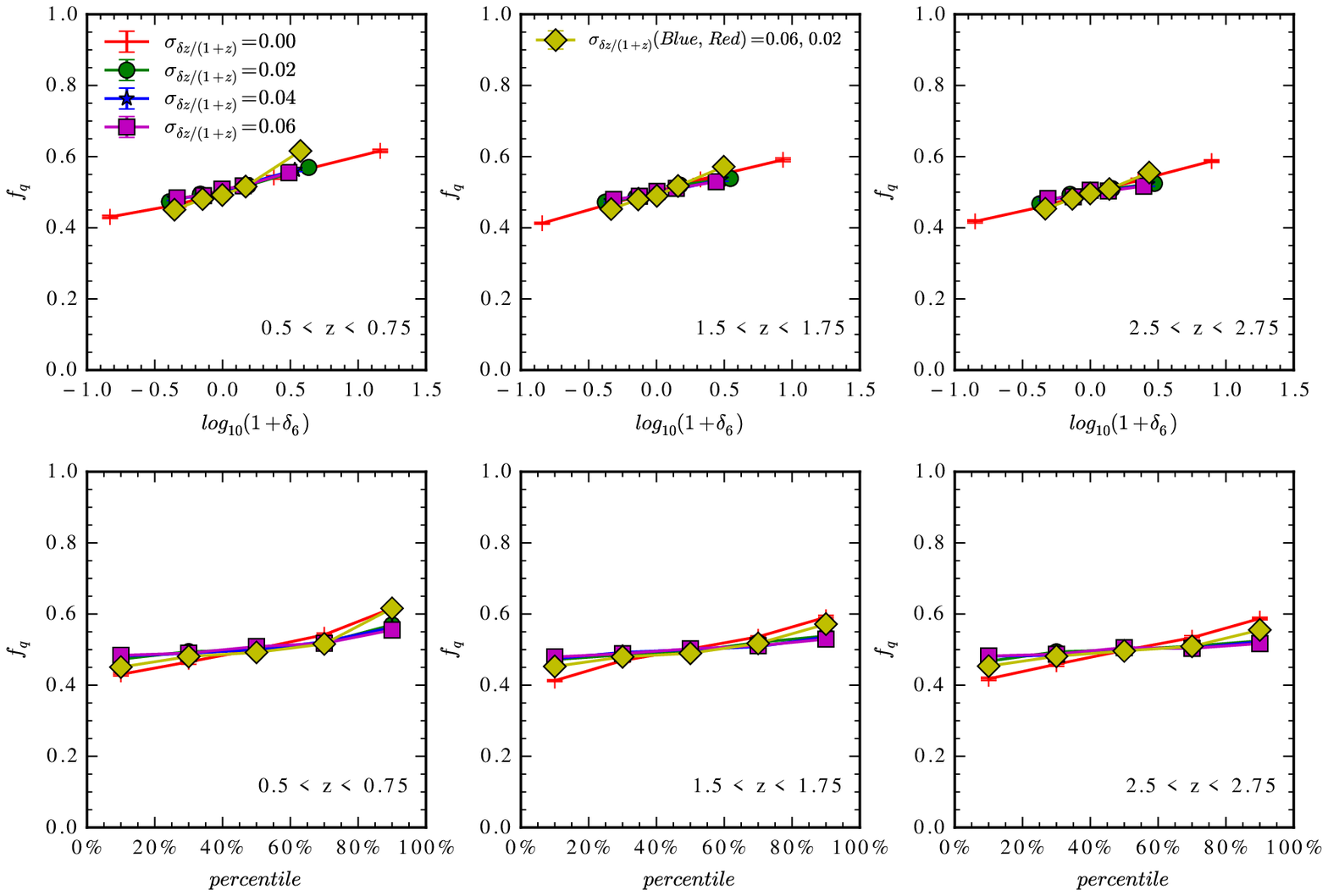}
\caption{Comparisons of quiescent fraction versus density between different photo-$z$ accuracies using the $6^{th}$-nearest neighbor method in a mock galaxy catalog. In the upper panels, the surface density field is presented using the overdensity, whereas in the lower panels the surface density field is shown using the percentage rank.
\label{fig:fq-photoz}}
\end{figure*}

\begin{figure*}
\includegraphics[angle=0,width=17cm]{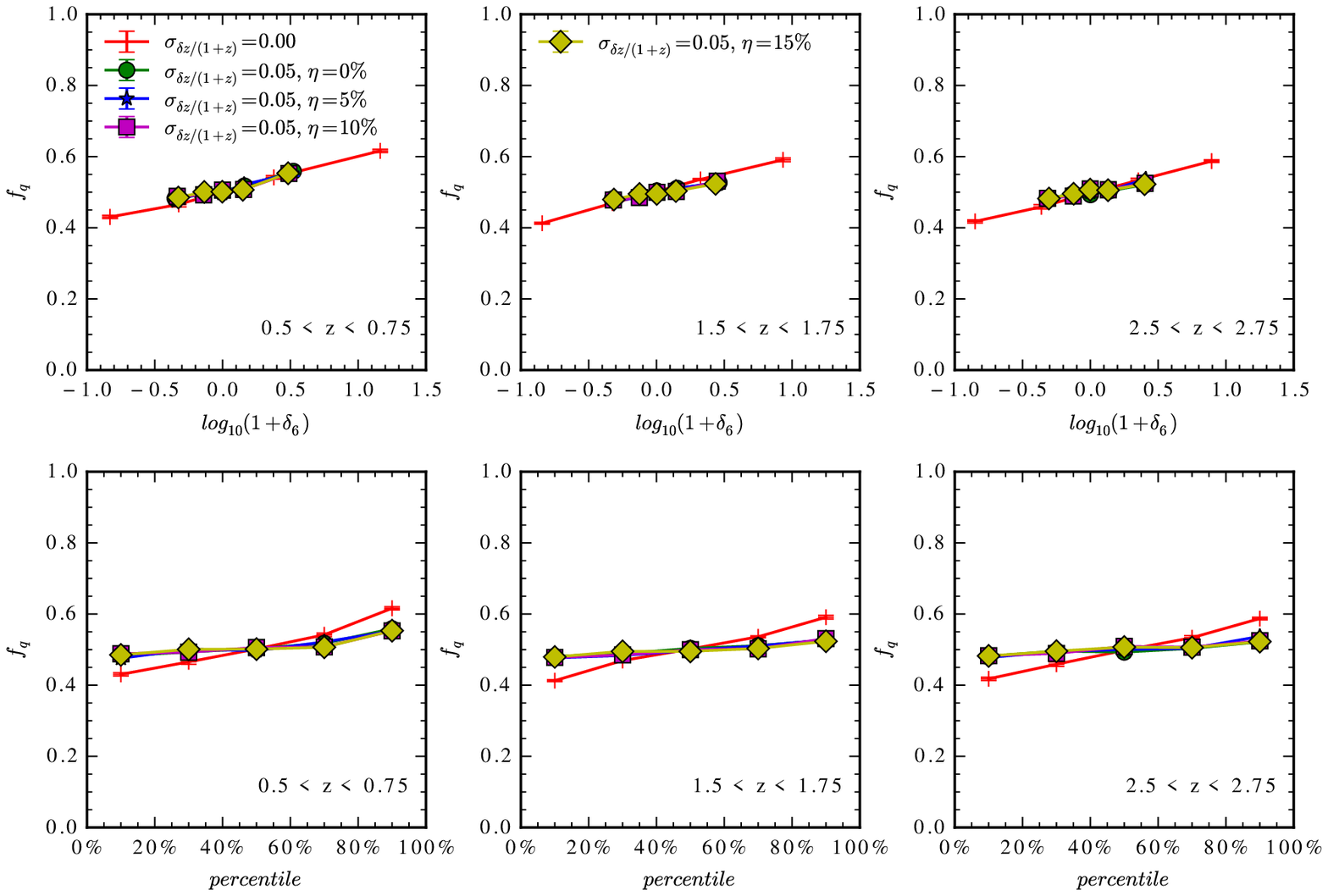}
\caption{Similar to figure \ref{fig:fq-photoz}, but the results are shown by varying the outlier rate. The photo-$z$ accuracy is fixed to 0.05 in all cases except for the magenta curves.
\label{fig:fq-outlier}}
\end{figure*}

Figures \ref{fig:fq-photoz} and \ref{fig:fq-outlier} display the quiescent fraction as a function of overdensity (upper panels) and density percentile (lower panels) under various photometric redshift precisions and fractions of catastrophic failure, respectively. 
We find that the resulting slopes of the \fq -- density relation only weakly depend on the redshift error and outlier rate. This is because even though the overdensity is diluted owing to the photo-$z$ error, which leads to a narrower overdensity range in the case of greater photo-$z$ uncertainty, the quiescent fraction for a given overdensity is also being diluted. As a result, the slope in the \fq --density relation remains nearly unaffected since values in both $x$- and $y$-axes lessen. Our test suggests that one can still recover the slope of the color--density relation with a photo-$z$ sample at least up to \deltaz~$\sim 0.06$. 

On the other hand, when the density is expressed using their percentage ranks rather than overdensity (as shown in the bottom panels of figures \ref{fig:fq-photoz} and \ref{fig:fq-outlier}), it is clear that the \fq -- density relation becomes much flatter in the presence of photo-$z$ errors. This is simply due to the fact that the range in the $y$-axis is reduced whereas the $x$-range does not change correspondingly. In other words, the adoption of the ranked density percentile could lead to the slope of color--density relation being degraded. Therefore, we choose to use the overdensity for our further analysis. Similarly, we found that the outlier rate has little effect on the color--density relation when using the overdensity, whereas the color--density relation is degraded when using the ranked density.

It is important to note that the aforementioned simulation tests are done based on a uniformly Gaussian scatter in the photometric redshift regardless of galaxy types, whereas in a realistic dataset the photometric redshift uncertainties for quiescent galaxies are typically smaller than star-forming galaxies because of their prominent spectral energy distribution features around the Balmer break. In order to know how this galaxy-type-dependent photometric redshift error can potentially affect our results, we repeat the above analysis by assigning photometric redshift to quiescent galaxies with the uncertainty 3 times better than the star-forming population. When compared to the mock sample with zero redshift errors, we find that there is a small excess ($<$ 20 \%) in the quiescent fraction in the densest regions, leading to a slightly higher slope of the \fq -- density (see figure \ref{fig:fq-photoz}). This means that our measured slope of the color--density relation can be seen as an upper limit when the slope is positive. Nevertheless, such an effect is small and therefore can be negligible in our analysis.

\subsection{Comparison of Spectroscopic and Photometric Redshift Density Measures}
\begin{figure*}
\includegraphics[angle=0,width=17cm]{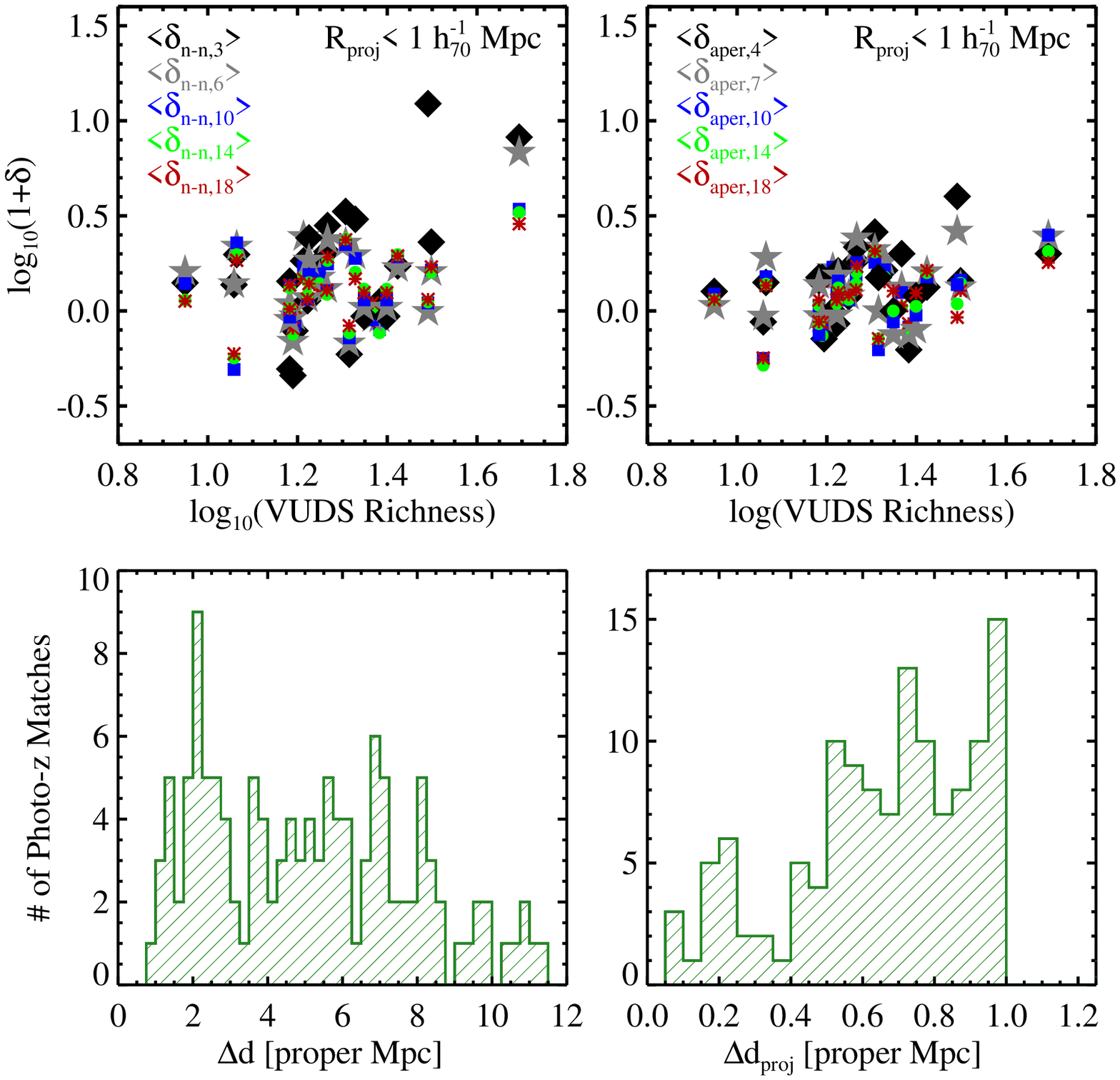}
\caption{\emph{Top Panels:} Plot of the logarithmic average photometric redshift density as measured by a variety of nearest neighbor 
(\emph{left}) and aperture (\emph{right}) measurements against a logarithmic spectroscopic estimate of density from proto-structures at 
$z_{spec}>1.5$ drawn from the VIMOS Ultra-Deep Survey (VUDS; Le F\`evre et al.\ 2015). Only those VUDS proto-structures that 
were found to be statistically significant are plotted. For each VUDS proto-structure, all photometric redshift density measurements 
with a median redshift within the redshift bounds of the proto-structure and a projected center within 1 $h_{70}^{-1}$ Mpc of the 
proto-structure center were averaged. Different metrics used to measure photometric redshift density are denoted by different 
symbols and colors ranging from small to larger scales (black filled diamonds and red asterisks, respectively). While there is 
an appreciable scatter both between the various metrics used to estimate $\log_{10}(1+\delta)$ and at a given $\delta_{\rm{VUDS}}$, 
a general and statistically significant trend of increasing $\log_{10}(1+\delta)$ with increasing $\delta_{\rm{VUDS}}$ is observed. 
\emph{Bottom Panels:} histogram of the offset of the matched $\log_{10}(1+\delta)$ estimates for all proto-clusters in real (\emph{left}) 
and projected (\emph{right}) space. For the former, differences along the line of sight were calculated from the median redshift of the 
galaxies in the matched photometric redshift density measurement and the median redshift of proto-structure members.}
\label{fig:VUDSvsphotdens}
\end{figure*}

We further test the reliability of our photometric redshift density measurements at higher density and higher redshifts ($z>1.5$).
In Figure \ref{fig:VUDSvsphotdens}, we plot $\log_{10}(1+\delta)$ against a proxy of density measurements of overdense 
environments in COSMOS drawn from the VIMOS Ultra-Deep Survey (VUDS; Le F\`evre et al.\ 2015), a massive 640 hour spectroscopic 
campaign targeting galaxies at $2\lsim z\lsim6$ across 1 $deg^2$ in three fields with the VIsible MultiObject Spectrograph 
(VIMOS; Le F{\`e}vre et al.\ 2003) mounted on the 8.2-m VLT at Cerro Paranal. For more details on the VUDS survey, we refer the reader to 
Le F\`evre et al.\ (2015). Spectroscopic density measurements were supplemented with additional redshifts from the 
$z$COSMOS faint survey (Lilly et al.\ \emph{in prep}). A search for spectroscopic overdensities (hereafter referred to as 
``proto-structures") in all three fields in VUDS was systematically performed over the redshift range $1.5 < z < 5$ roughly
following the method of \citet{lem14}. In total, 26 statically significant proto-structures were discovered in the 
area of the COSMOS field covered by VUDS over the redshift range $1.6 < z < 4.5$ of which one has already been reported 
(Cucciati et al.\ 2014). 

We plot in the top panels of Figure 7 the measurement of the strength of the overdensity of each VUDS proto-structure (on the abscissa), defined as the number of spectroscopic members divided by the `filter' size encompassing the members normalized to the maximal filter size defined in \citet{lem14}. This quantity, $\delta_{VUDS}$, is equal to
$\delta_{VUDS}\sim0.3$ for the average region in VUDS and roughly scales with $\delta_{gal}$; thus, it may be used as a proxy of both the significance and halo mass of the proto-structure. For each VUDS proto-structure, all photometric
density measurements based on a given metric with a median redshift within the proto-structure and a projected center 
coincident within $R_{proj}\leq1.0$  $h_{70}^{-1}$ Mpc of the proto-structure were averaged. All 24 VUDS/COSMOS proto-structures with 
$z<3.5$ had at least one photometric redshift density measurement match, with the vast majority having several matches. 
This exercise was repeated for a variety of different nearest neighbor and aperture photometric redshift density 
measurements ranging from the smallest scales probed in this study ($\delta_{n-n,3}$ and $\delta_{aper,4}$) to the largest
($\delta_{n-n,18}$ and $\delta_{aper,18}$). While a general trend of increasing $\log_{10}(1+\delta)$ with $\delta_{\rm{VUDS}}$ 
is observed in the top panels of Figure \ref{fig:VUDSvsphotdens}, there exists a large amount of scatter. Thus, for each 
photometric density measurement metric, we performed a Spearman's test to estimate the significance of correlation between
our measurements and those from VUDS. For individual photometric redshift density measurements the significance of the correlation
ranged between $\sim0.5\sigma$ and $\sim2\sigma$, with generally more significant correlation on larger scales, perhaps not
surprising given the tendency of overdensities at these redshifts to be extended (see, e.g., Chiang et al.\ 2013). However, 
when combining all photometric density measurements of a given method, we find a significant correlation ($\sim3\sigma$) between
both the aperture and nearest neighbor photometric redshift density measurements with $\delta_{VUDS}$. While there exists significant
uncertainty in the photometric redshift density for a given measurement and for a particular proto-structure, the observed 
concordance with spectroscopic overdensity measurements likely means that we are, on average, correctly identifying overdense 
environments.

\section{Quiescent Fraction as a Function of Local Density} \label{sec:fq}

\begin{figure*}
\includegraphics[angle=-270,width=17cm]{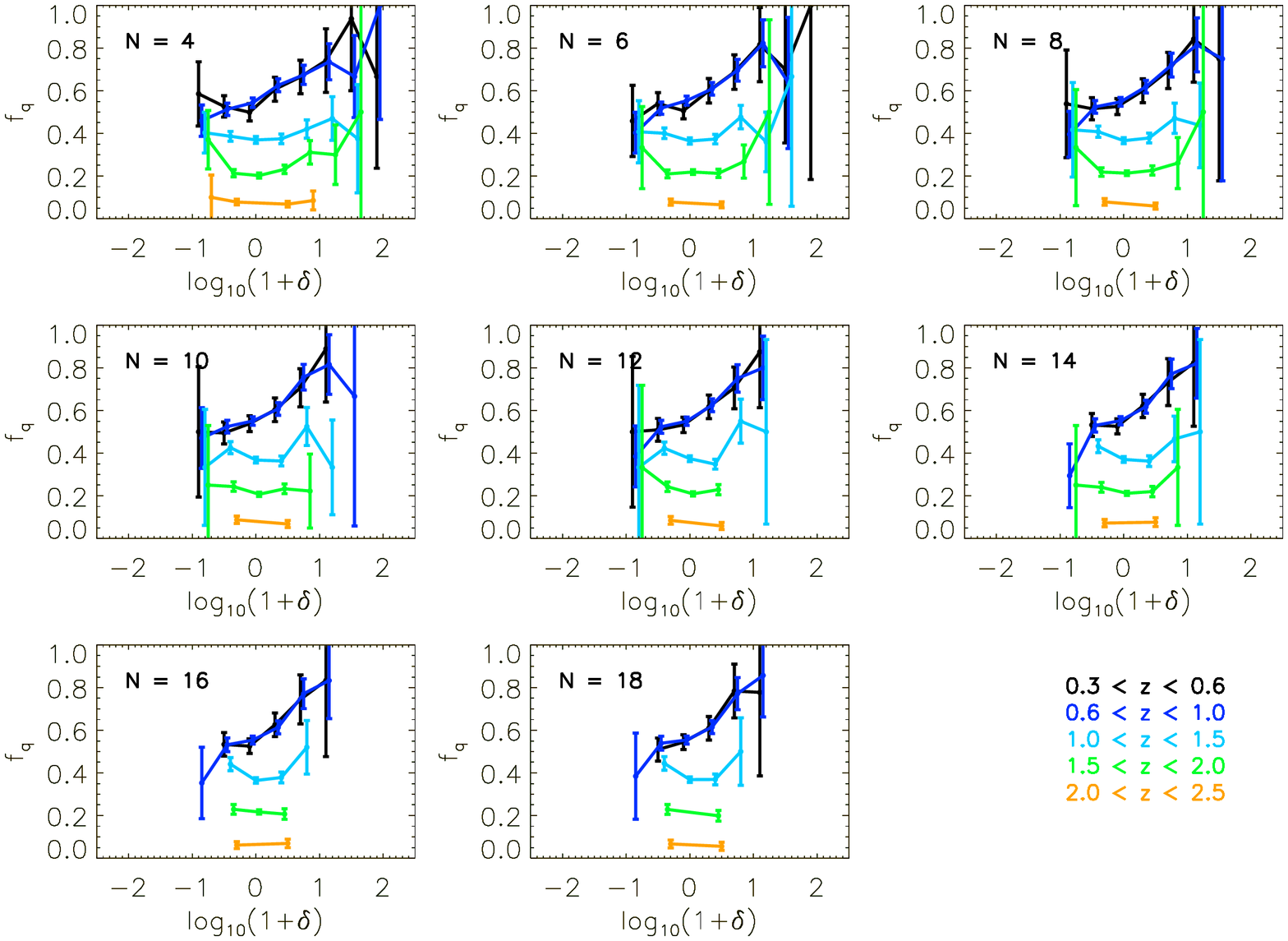}
\caption{Quiescent fraction as a function of overdensity for five redshift intervals (black: $0.3 < z < 0.6$; blue: $0.6 < z < 1.0$; cyan: $1.0 < z < 1.5$; green: $1.5 < z < 2.0$; orange: $2.0 < z < 2.5$). Each panel corresponds to a choice of the scale of nearest neighbor used for the density measurement. The data points are shifted slightly in the x-axis for clarity.
\label{fig:fq}}
\end{figure*}

\begin{figure*}
\includegraphics[angle=-270,width=17cm]{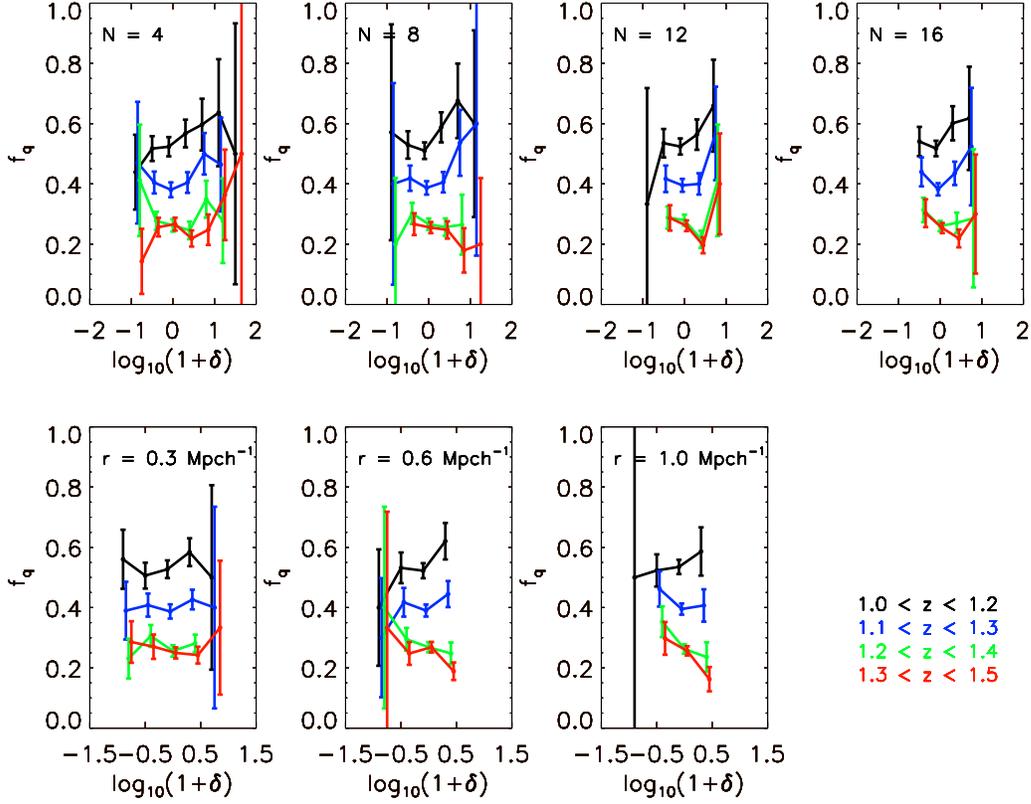}
\caption{Zoomed-in view of the quiescent fraction versus overdensity at $1.0 < z < 1.5$ using both the nearest neighbor (top panels) and the fixed aperture (bottom panels) methods. The aperture size (in units of \mpc~ in physical coordinates) or the choice of $N$ increases from left to right. 
\label{fig:fq_narrowz}}
\end{figure*}

\begin{figure}
\includegraphics[angle=270,width=8.5cm]{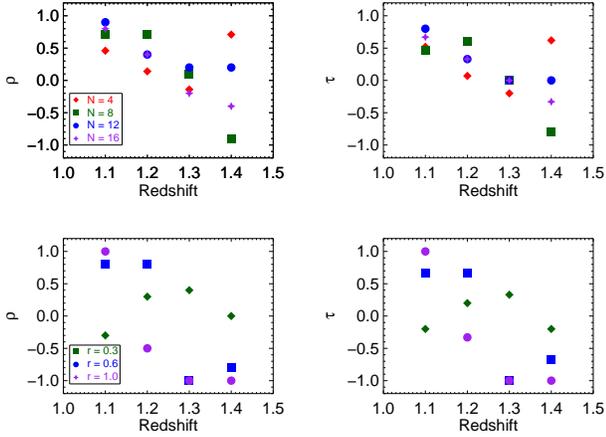}
\caption{\emph{Top Panels:} Spearman's (\emph{left}) correlation coefficient ($\rho$) and Kendall's (\emph{right}) correlation coefficient ($\tau$) for the \fq--density relation in the range of $1 < z <1.5$ with a variety of nearest neighbor measurements. \emph{Bottom Panels:} same as top panels, but measured by using the aperture method with various projected aperture sizes.
\label{fig:correlation}}
\end{figure}

\begin{deluxetable}{llcccc}
\tabletypesize{\scriptsize}
\tablewidth{0pt}
\tablecaption{Correlation coefficients for the \fq--density relation.\label{tab:correlation}}
\tablehead{
    \colhead{Subsample} &
    \colhead{Redshift} &
    \colhead{$\rho^{a}$} &
    \colhead{$S_{\rho}^{b}$} &
    \colhead{$\tau^{c}$} &
    \colhead{$S_{\tau}^{d}$}
}

\startdata
               N = 4 &  1.0 < z < 1.2 &  0.46 &  0.29 &  0.52 &  0.10\\
               N = 4 &  1.1 < z < 1.3 &  0.14 &  0.79 &  0.07 &  0.85\\
               N = 4 &  1.2 < z < 1.4 & -0.14 &  0.79 & -0.20 &  0.57\\
               N = 4 &  1.3 < z < 1.5 &  0.71 &  0.07 &  0.62 &  0.05\\
\hline
               N = 8 &  1.0 < z < 1.2 &  0.71 &  0.11 &  0.47 &  0.19\\
               N = 8 &  1.1 < z < 1.3 &  0.71 &  0.11 &  0.60 &  0.09\\
               N = 8 &  1.2 < z < 1.4 &  0.10 &  0.87 &  0.00 &  1.00\\
               N = 8 &  1.3 < z < 1.5 & -0.90 &  0.04 & -0.80 &  0.05\\
\hline
              N = 12 &  1.0 < z < 1.2 &  0.90 &  0.04 &  0.80 &  0.05\\
              N = 12 &  1.1 < z < 1.3 &  0.40 &  0.60 &  0.33 &  0.50\\
              N = 12 &  1.2 < z < 1.4 &  0.20 &  0.80 &  0.00 &  1.00\\
              N = 12 &  1.3 < z < 1.5 &  0.20 &  0.80 &  0.00 &  1.00\\
\hline
              N = 16 &  1.0 < z < 1.2 &  0.80 &  0.20 &  0.67 &  0.17\\
              N = 16 &  1.1 < z < 1.3 &  0.40 &  0.60 &  0.33 &  0.50\\
              N = 16 &  1.2 < z < 1.4 & -0.20 &  0.80 &  0.00 &  1.00\\
              N = 16 &  1.3 < z < 1.5 & -0.40 &  0.60 & -0.33 &  0.50\\
\hline
 r = 0.3 $h^{-1}$Mpc &  1.0 < z < 1.2 & -0.30 &  0.62 & -0.20 &  0.62\\
 r = 0.3 $h^{-1}$Mpc &  1.1 < z < 1.3 &  0.30 &  0.62 &  0.20 &  0.62\\
 r = 0.3 $h^{-1}$Mpc &  1.2 < z < 1.4 &  0.40 &  0.60 &  0.33 &  0.50\\
 r = 0.3 $h^{-1}$Mpc &  1.3 < z < 1.5 &  0.00 &  1.00 & -0.20 &  0.62\\
\hline
 r = 0.6 $h^{-1}$Mpc &  1.0 < z < 1.2 &  0.80 &  0.20 &  0.67 &  0.17\\
 r = 0.6 $h^{-1}$Mpc &  1.1 < z < 1.3 &  0.80 &  0.20 &  0.67 &  0.17\\
 r = 0.6 $h^{-1}$Mpc &  1.2 < z < 1.4 & -1.00 &  0.00 & -1.00 &  0.04\\
 r = 0.6 $h^{-1}$Mpc &  1.3 < z < 1.5 & -0.80 &  0.20 & -0.67 &  0.17\\
\hline
 r = 1.0 $h^{-1}$Mpc &  1.0 < z < 1.2 &  1.00 &  0.00 &  1.00 &  0.04\\
 r = 1.0 $h^{-1}$Mpc &  1.1 < z < 1.3 & -0.50 &  0.67 & -0.33 &  0.60\\
 r = 1.0 $h^{-1}$Mpc &  1.2 < z < 1.4 & -1.00 &  0.00 & -1.00 &  0.12\\
 r = 1.0 $h^{-1}$Mpc &  1.3 < z < 1.5 & -1.00 &  0.00 & -1.00 &  0.12
\enddata

\tablecomments{${(a)}$ \& ${(b)}$: These two column denote the Spearman's rank correlation coefficient and the significance of its deviation from zero ; ${(c)}$ \& ${(d)}$: These two column denote the Kendall's rank correlation coefficient and the significance of its deviation from zero.}

\end{deluxetable}

\begin{figure*}               
\centering
\includegraphics[angle=-270,width=17cm]{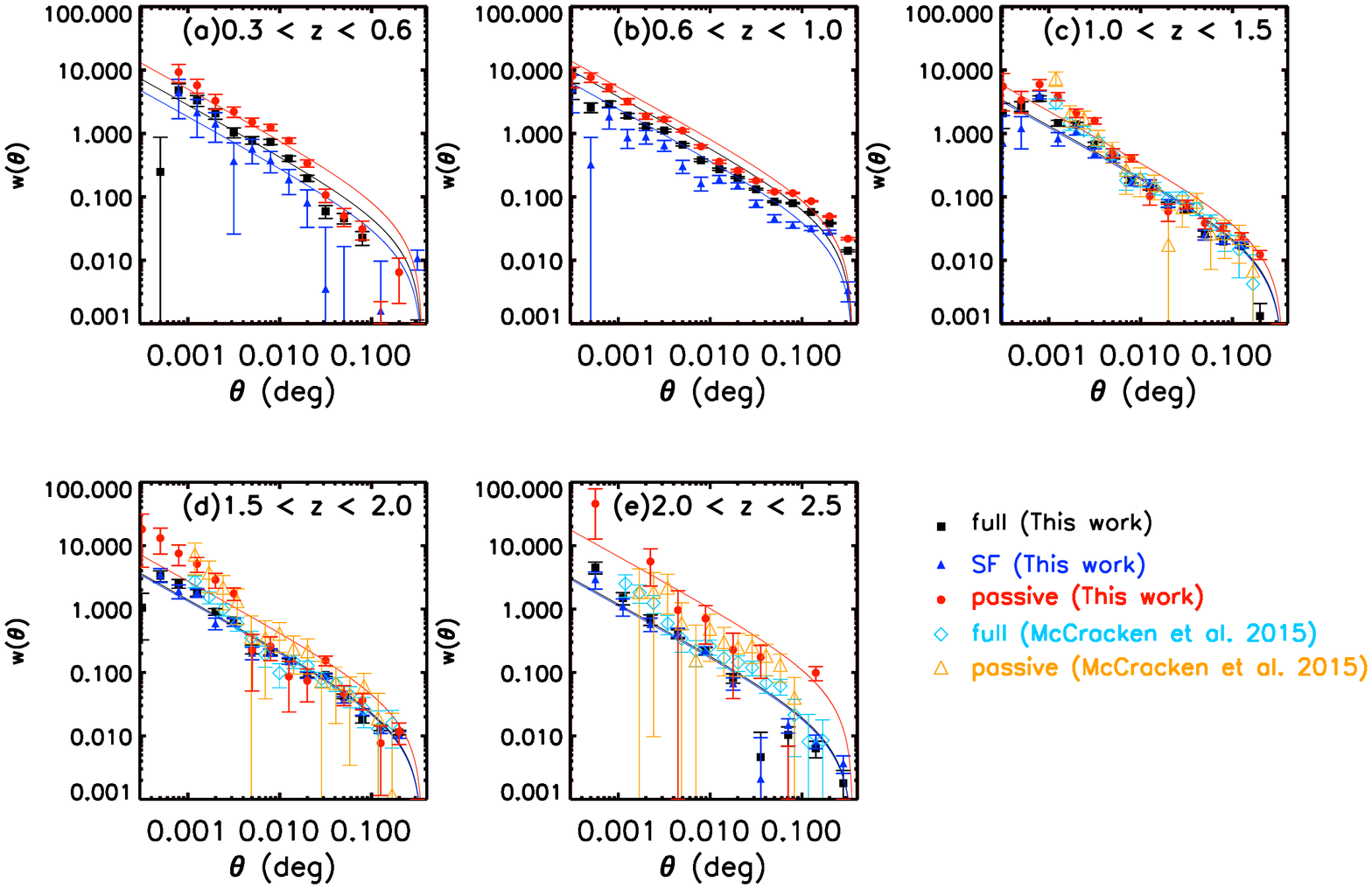}
\caption{ACF for the full (black squares), star-forming (blue triangles), and quiescent (red circles) populations from $z = 0.3$ to $z = 2.5$. We also overplot the results of \citet{mc15} for the full sample (open cyan diamonds) and the passive galaxy sample (open orange triangles) with \sm~ $> 10^{10.6}$ \Msolar.
  \label{fig:acf}}
\end{figure*}

After constructing the density field, we now present the quiescent fraction, defined as the ratio of the number of quiescent galaxies to that of the full sample, as a function of local density. Figure \ref{fig:fq} shows the quiescent fraction for galaxies with \sm $> 10^{10.5}$ \Msolar~ from $z = 0.3$ up to $z \sim 2.5$.  At $z < 1$, the quiescent fraction strongly depends on the overdensity, being larger in denser environments, regardless of the scales used to measure the density field. This is in agreement with previous results in the literature \citep{coo08,sco13,hah15}. At higher redshifts, we observe that the quiescent fraction versus density relation flattens or even inverts at $z > 1$. 

For further inspection, we divide the sample into narrower redshift bins as shown in figure \ref{fig:fq_narrowz}. The upper and lower panels display the results using the nearest neighbor and aperture methods, respectively. For a fixed environment scale, there exists a general trend that the \fq --density relation  changes from positive to negative with increasing redshift, except on very small scales (small `$N$' or `$r$').  This trend is more prominent when the aperture method is applied. The turning redshift point is $\sim 1.2$ or so. To further quantify the correlation between quiescent fraction and density field, we compute both the Spearman's and Kendall's rank correlation coefficients in each case. The results of the correlation and the significance of its deviation from zero are shown in figure \ref{fig:correlation} and in Table \ref{tab:correlation}. However, the significance of the negative slope varies from method to method (aperture versus nearest neighbor) and is up to 1.8 $\sigma$ at most. For example, when considering $r$ = 0.6 and 1.0 \mpc, the reversal of the slope with respect to redshift is only a 1.4 and 1.7 result, respectively.

Furthermore, there is marginal evidence (<1.8$\sigma$) that the slope of the \fq--density relation is scale dependent in the two higher-redshift bins ($1.2 < z < 1.4$ and $1.3 < z < 1.5$), decreasing with increasing `$N$' or `$r$'. The Kolmogrov--Smirnov (K-S) test also suggests that the probability that the large-scale density distributions for the star-forming and quiescent populations are drawn from the same parent distribution varies from 0.1\% to 9\%, indicating a marginal difference in the two density distributions when using large `$N$' or `$r$'.  The negative correlation coefficients seen at larger-scale environments suggests that galaxies located in denser large-scale environments are less efficient in shutting off their star formation.

It has been noted that the color--density (or specific SFR [SSFR]--density) relation has been highly debated in the literature at $z \sim 1$. By using a photometric reshift sample in the GOODS fields, there is a claim that SSFR is anticorrelated with density at $z\sim 1$ \citep{elb07}, while other works have concluded that the color--density relation persists out to $z \sim 2$ \citep{coo08,qua12}. On the other hand, there are also studies showing a very weak environmental dependence of the SSFR \citep{sco13}.  
Direct comparison among the different works employing different density measurements (aperture versus nearest neighbor), redshift accuracy (spectroscopic versus photometric), and sample depth is very complicated and beyond the scope of this work.
Nevertheless, our work suggests that the discrepancy among different studies may be partly attributed to the redshift and scale dependence of the color-density relation. As we have discussed, the turning point of the color--density relation appears at redshifts between 1 and 1.5. As a consequence, the slope of the color--density relation would be sensitive to the redshift range probed. Since the galaxy sample of \citet{coo08} is restricted to $z < 1.0$, as opposed to $z < 1.2$ adopted by \citet{elb07}, the reversal of the color--density relation is naturally not seen by \citet{coo08}. Furthermore, in the work of \citet{coo08}, the authors utilized spectroscopic redshifts and computed the overdensity using the $3^{rd}$-nearest neighbor approach. This roughly corresponds to a median size of $\sim$0.5 physical \mpc~ at $z\sim1$.  
On the other hand, despite that \citet{elb07} also adopted a comparable aperture size of 0.5 physical \mpc~  on the projected plane, they need to adopt a larger window in the redshift direction because of the photometric redshift uncertainty; their density fields are in fact measured on a much larger scale. The combination of both effects can potentially explain the different results obtained by these two groups.

When moving to redshifts greater than 1.5, our derived quiescent fraction depends little on the density field, consistent with other recent works that are also based on the COSMOS sample \citep{sco13,dar15}. However, there is also tentative evidence that the \fq--density relation stays inverted out to $z\sim2.5$, although the dynamical range of the density at high redshifts is smaller compared to $z < 1.5$. Nevertheless, we stress that the current sample size and the survey area are  still small. Observations over a larger field of view or in other fields are required to confirm the features seen here.

\section{Clustering as a Function of Galaxy Properties\label{sec:clustering}}
While the local density field has been widely used to explore the effect of the environment on galaxy properties, it does not represent the overarching environments such as voids, filaments, groups, and clusters with a one-to-one correspondence. For instance, the local density in a certain area of a filament can be as high as that within a galaxy group, which makes it difficult to interpret the effect of the environment in terms of actual physical processes. In the previous section, we have probed the dependence of the quiescent fraction on local density. Now we turn to the study of the dependence on physical environment (or halo mass). Ideally it is preferable to associate each galaxy with their underlying physical environments and then compare galaxy properties among different environments. Such an approach, however, requires high redshift accuracy for reliable environment constructions. As an alternative, we perform a clustering analysis of star-forming and quiescent galaxies to probe the hosting halo mass for the two populations. If the galaxy density field is indeed a good tracer of halo mass, one might well expect the clustering strength for quiescent galaxies to be larger than that of star-forming galaxies at $z < 1$ as inferred from the \fq--density relation. At $1 < z <1.5$, where the \fq--density relation is inverted at large scales, SF galaxies are expected to have a stronger clustering length than quiescent galaxies do.

We measure the galaxy angular correlation function (ACF) using data and random catalogs with the estimator proposed by \citet{lan93}:
\begin{equation}\label{eq_ac}
\omega(\theta) = [DD(\theta)(n_{R}/n_{D})^{2}-2DR(\theta)(n_{R}/n_{D})+RR(\theta)]/RR(\theta),
\end{equation}
where DD($\theta$), DR($\theta$), and RR($\theta$) are the number of data--data, data--random and random--random pair counts with separations between $\theta$ and $\theta+\delta \theta$, respectively, and $n_{D}$ and $n_{R}$ are number of galaxies in the data and random catalogs, respectively. A mask is used to identify bad regions due to saturated stars or cosmic rays and is applied to both the data and random catalogs. The error bars in each $\theta$ bin are based on the 1$\sigma$ Poisson statistics of the pair counts in the bins and hence are not correlated across the bins. We fit our ACF with a power law
\begin{equation}
\omega(\theta) = A_{\omega}(\theta^{1-\gamma} - C),
\end{equation}
where $\gamma$ is fixed to be 1.8 and $C$ is the integral constraint that accounts for the finite region of the sky probed in the sample. Following \citet{roc99}, we estimate the integral constraint using
\begin{equation}\label{eq_Cint}
C=\frac{\Sigma RR(\theta)\theta^{-0.8}}{\Sigma RR(\theta)},
\end{equation}
which gives a value of $C = 2.28$ in our case.

In figure \ref{fig:acf}, we show the ACF for SF (blue points), quiescent (red points), and the full (black points) samples. Quiescent galaxies persistently have a greater clustering amplitude than SF galaxies up to $z\sim2.5$, indicating that quiescent galaxies preferentially live in more massive halos compared to the SF population. Our clustering results for the full and quiescent samples are in good agreement with those from an independent analysis based on the UltraVISTA DR1 data in the COSMOS field \citep{mc15}, which also found that quiescent galaxies are more strongly clustered than the full sample. However, we note that the data used in this work incorporate an updated version (DR2) of the UltraVISTA NIR survey and SPLASH IRAC survey, which yield more reliable mass estimates compared to previous photometric redshift catalogs in this field. At $z < 1$, galaxy type dependence of the clustering is consistent with the results from the density analysis shown in \S \ref{sec:fq}: a high-density region provides a higher chance of finding quiescent galaxies. 

In contrast, the clustering results seem to disagree with the expectation from the \fq--density relation at higher redshifts--at $z > 1.5$ quiescent galaxies are clustered more strongly, while the \fq~ has little dependence or is inversely correlated with density. In other words, our results indicate that even if the typical halo mass of quiescent galaxies is generally higher than that of SF galaxies at all redshifts, the galaxy-based density for quiescent galaxies is not necessarily higher. 

\begin{figure*}
  \centering
  \includegraphics[angle=0,width=15cm]{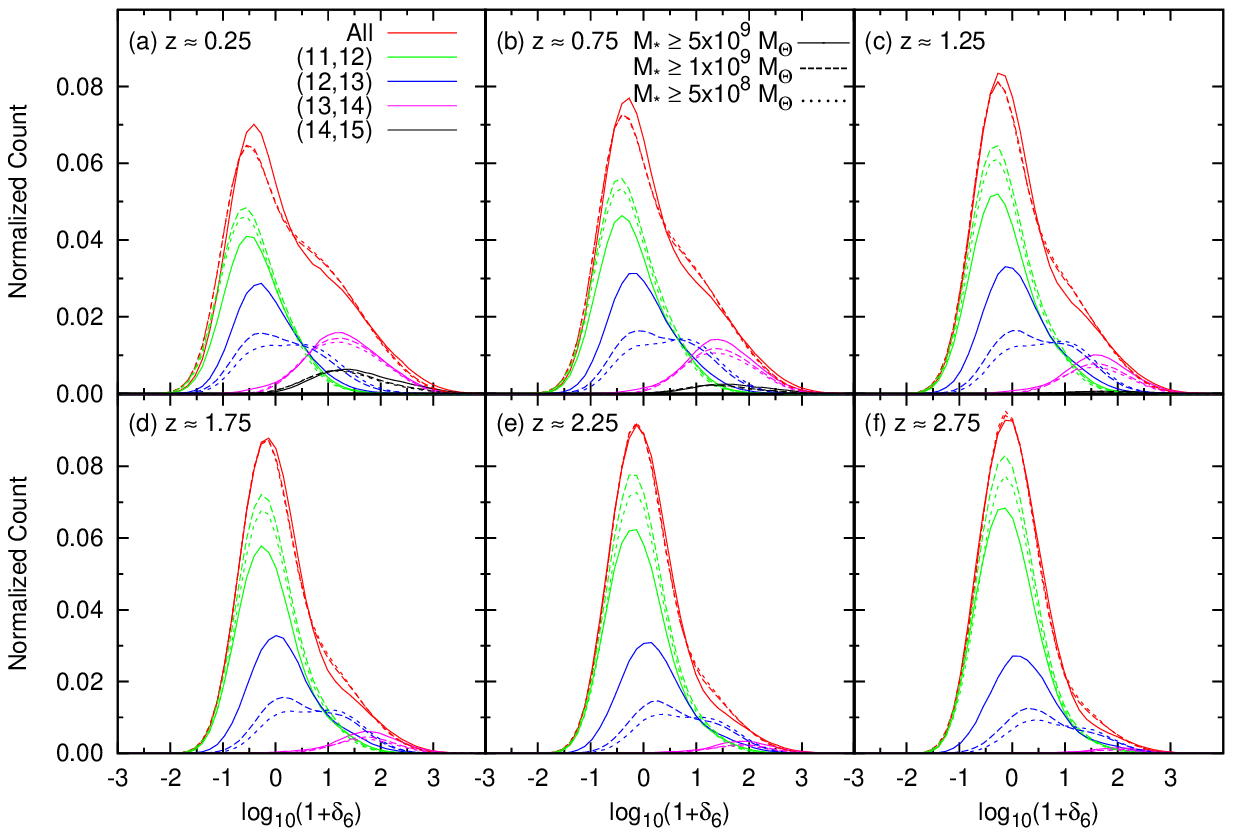}
  \caption{Distributions of overdensity  for galaxies located in halos with various mass ranges (red: all; green: $10^{11} < M_{halo}$/\Msolar $< 10^{12}$; blue: $10^{12} < M_{halo}$/\Msolar $< 10^{13}$; magenta: $10^{13} < M_{halo}$/\Msolar$ < 10^{14}$; black: $10^{14} < M_{halo}$/\Msolar$ < 10^{15}$). Solid, dashed, and dotted lines represent the galaxy local density measured using neighboring galaxies more massive than $5\times10^{9}$, $10^{9}$ and $5\times10^{8}$ \Msolar, respectively.\label{fig:mock}}
\end{figure*}

\begin{figure}
  \centering
  \includegraphics[angle=-270,width=10cm]{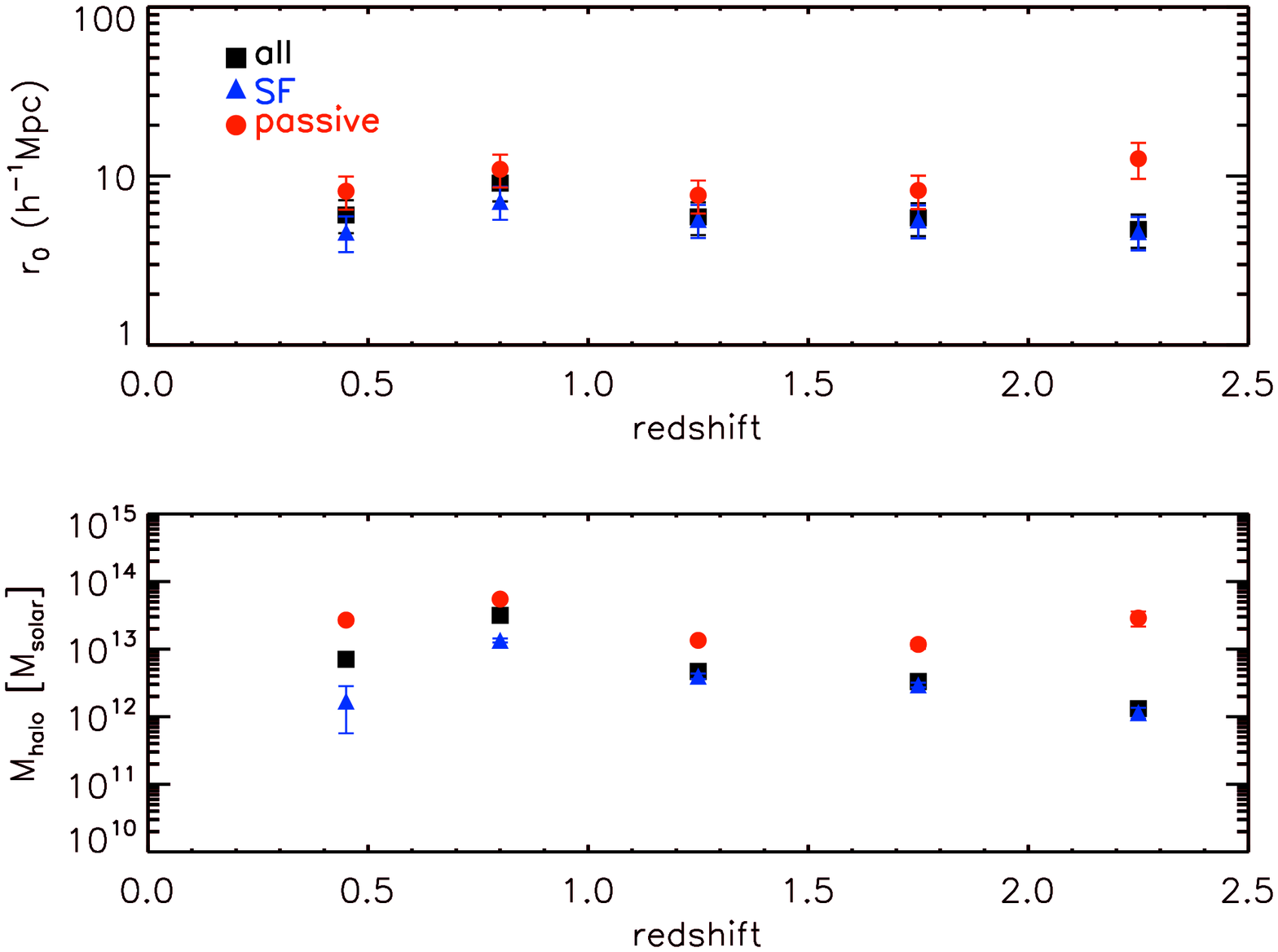}
  \caption{Redshift evolution of the correlation length (upper panel) and the corresponding halo mass (lower panel) for the full (black squares), star-forming (blue triangles), and quiescent (red circles) populations.
  \label{fig:r0}}
\end{figure}

In fact, it has been demonstrated that lower halo masses, in the regime 10$^{11}$-10$^{13}$ \Msolar, are indiscernible by various density measurements, including the nearest neighbor and aperture methods \citep{mul12,lai15}. To have a closer look at the relationship between halo mass and galaxy density, we make use of the mock galaxy catalog and plot their distributions in figure \ref{fig:mock}. Here we fix the stellar mass to be $>5\times10^{8}$, $10^{9}$ and $5\times10^{9}$ \Msolar, respectively, when measuring the overdensity inferred from the $6^{th}$-nearest neighbor. As can be seen, although the peak of the density distribution is in general higher for a larger halo mass bin, there is a wide spread in density for a given halo mass bin. Moreover, the density distribution becomes degenerate for halos with mass less than $10^{13}$~\Msolar, confirming the results found by other works \citep{mul12,lai15} using different simulations. 
As we will discuss in the next paragraph (also see figure \ref{fig:r0}), we find that the typical halo mass at $z>2$ is roughly $10^{13}$ \Msolar~ or so, a mass regime below which the local density is no longer a good tracer of halo mass. The large scatters in the relationship between local density and halo mass may explain the disagreement between the density and clustering analysis at high redshifts.

Figure \ref{fig:r0} illustrates the redshift evolution of the correlation length $r_{0}$ and the inferred mass of the hosting halo based on the \citet{mo02} prescription for our sample with \sm $> 10^{10.5}$ \Msolar. The star-forming galaxies show an increase in the halo mass with time (except for the lowest-$z$ bin). This trend can possibly be explained by the global growth of the large-scale structures with time, as fractionally more galaxies become part of the galaxy groups and clusters as time progresses. On the other hand, the evolution in halo mass is much weaker for the quiescent population. However, we note that the expected number of clusters more massive than $10^{14}$ \Msolar~ at $z < 0.5$ in a COSMOS-like field is $\sim4$. It is likely that our sample at $z < 0.5$ does not sample enough volume to detect massive halos, resulting in the decline of the typical halo mass for both star-forming and quiescent galaxies from $z \sim 0.8$ to $z \sim 0.5$, as revealed in figure \ref{fig:r0}. The weaker evolution for the halos hosting a quiescent population may therefore be partly attributed to the lack of volume at $z < 0.5$. At $z \sim 2-2.5$, the typical halo mass for quiescent galaxies is $10^{13}$ \Msolar, at least one order of magnitude more massive than that for the SF galaxies. This implies that the quenching mechanism acting in the high-redshift Universe may only be effective in massive halos.

\section{DISCUSSION AND CONCLUSIONS}\label{sec:discussion}

Using the SPLASH photometric redshift catalog in the COSMOS field, we have investigated the environmental effects on galaxy properties by studying the density field and clustering properties of star-forming and quiescent galaxies out to $z \sim 2.5$. We first test the performance of the density recovery with a photometric-redshift-like catalog by using a mock galaxy catalog, and we study the impact of the photometric redshift uncertainty on the color--density relation. We then use our density measurements to probe the fraction of quiescent galaxies as a function of density with various scales. Finally, we perform clustering analyses for both star-forming and quiescent populations to infer their hosting halo masses.
In summary, the key results from our analysis are as follows:

1. Extensive simulations show that the color--density relation can still be detectable even if the surface density field is diluted in the presence of photometric redshift errors and catastrophic failures. However, the slope of the color--density  relation depends little on redshift uncertainty when using the overdensity, but would be weakened if using the ranked percentage of density.

2. We find that the fraction of quiescent galaxies (\fq) is a strong function of density field at $z < 1$, being greater in denser environments. However, the \fq --density relation flattens and becomes inverted between $z = 1.0$ and $1.5$ at a 1.6$\sigma$ confidence level. 
There is also marginal evidence (<1.8$\sigma$) that the slope of the \fq --density relation is scale dependent at $1 < z < 1.5$, changing from positive to negative with increasing environment scale.

3. Beyond $z =1.5$, the quiescent fraction depends little on their density fields. There is very weak evidence (<1.4$\sigma$) that denser environments show smaller quiescent fraction.

4. For massive  (\sm~ > 10$^{10.5}$ \Msolar) galaxies, the quiescent population possesses greater correlation lengths than the star-forming population at $z = 0.3-2.5$, indicating that quiescent galaxies preferentially reside in more massive halos compared to star-forming galaxies at a fixed stellar mass. 

5. The lack of density dependence of the quiescent fraction, together with the clear difference in the clustering strength (and thus halo mass) seen between quiescent and star-forming galaxies, suggests that halo mass is a more fundamental parameter in suppressing the star formation in galaxies, as opposed to the density field in the high-redshift Universe.

One interesting result from this work is that not only does the \fq--density relation reverse its trend at $z \sim 1$, but the slope of this relation is also scale dependent, inverted on larger scales when adopting finer redshift bins in the redshift range of $1 < z < 1.5$. The higher passive fraction seen in large-scale underdense environments  at $z \sim 1$ is somewhat unexpected and not predicted in galaxy formation models \citep{elb07}, particularly because the quiescent fraction is sensitive to the quenching processes, which are intuitively thought to be more effective in high-density environments (e.g., galaxy mergers, ram pressure stripping, strangulations). We note, however, that the cessation of star formation requires cold gas to be completely removed, consumed, or heated to a higher temperature. If new cold gas can be replenished into the galaxy after quenching occurred, the galaxy cannot remain quiescent forever. Galaxies located in the denser part of the cosmic web where cold gas is more abundant may regain their gas through the cold accretion mode more easily. On the other hand, the positive \fq-density relation seen on small scales implies that the quenching process subject to the local environment similar to the local Universe is still active at $z \sim 1$. In other words, there are two competing environmental effects for galaxies located in high-density regions, where both quenching and the revival of star formation are more efficient. However, we stress that the reversal of the \fq-density relation is only detected at a 1.5$\sigma$ level. A larger dataset would be required to improve the statistics.

At $z> 1.5$ there is no significant density dependence of the quiescent fraction, in contrast to the halo mass dependence. As discussed in \S5, this can be mostly attributed to the fact that at high redshifts, most of the massive relaxed clusters whose density fields are distinct from those of smaller halos had not had enough time to develop, and that there is fairly large scatter in the relationship between the density field and halo mass for less massive halos (<$10^{13}$\Msolar). In other words, there is only weak one-to-one correspondence between the density field and halo mass for our high-redshift sample. Furthermore, even if the extreme but rare environments do exist, they either may be missing, owing to the small field size used in this work, or are at different stages of virialization from their local counterpart. Therefore, the environment effects we see for the high-redshift sample may exclude the quenching processes that only operate in very massive clusters, such as those seen at low redshifts.

Theoretical works predict that halos more massive than $10^{12}$\Msolar~ can form a hot gaseous halo via virial shock \citep{bir03,ker05,dek06}. In the presence of heating sources, such as active galactic nucleus (AGN) or radio-mode feedback, the hot gas fails to cool efficiently and thus prevents further star formation in galaxies. This type of quenching mechanism is often referred to as `halo quenching'. Observationally, it has also been suggested that halo mass is a primary
parameter responsible for galaxy quenching (e.g., Zu \& Mandelbaum 2015).
Under this paradigm, it has been shown that the hot gas fraction increases with halo mass \citep{gab15}. If we believe star formation quenching efficiency to be directly related to the hot gas fraction, it is then expected that the quiescent fraction of galaxies would also be a strong function of halo mass. In this study, we show that at $z \sim 1-2.5$, the typical halo mass hosting the quiescent galaxies is $10^{13}$ \Msolar, relatively higher compared to that of star-forming galaxies, suggesting that massive halos are more likely to provide conditions that maintain the quenched state of galaxies. This is in line with the aforementioned halo quenching picture, in which the quenching efficiency is higher for more massive halos.

We caution that the number of quiescent galaxies at higher redshifts in our sample is still very limited, which leads to large statistical error bars especially in extreme environments. Wider and deeper surveys such as the Hyper-Suprime-Cam Survey (HSC) combined with the SPLASH program will increase the sample size of quiescent galaxies and push the analysis to fainter populations. The increased statistics will in turn allow the hosting halo mass to be better constrained by the HOD analysis. Furthermore, the upcoming large spectroscopic surveys, such as Prime Focus Survey (PFS) on $Subaru$, will allow us to construct the cosmic web by identifying a broad range of environments (voids, filaments, groups, and clusters) and provide direct evidence for the environmental effects in the high-redshift Universe.

\acknowledgments

We thank the anonymous referee for constructive suggestions that significantly improved the clarity of this paper. We also thank C.-C. Lai for useful discussions. The work is supported by the Ministry of Science and Technology (MoST) in Taiwan under the grant MOST 103-2112-M-001-031-MY3. C.L. is supported by the ILP LABEX (ANR-10-LABX-63 and ANR- 11-IDEX-0004-02). The UltraVISTA data are based on data products from observations made with ESO Telescopes at the La Silla Paranal Observatory under ESO program ID 179.A-2005 and on data products produced by TERAPIX and the Cambridge Astronomy Survey Unit on behalf of the UltraVISTA consortium.

\end{document}